\documentclass[aps,prb,preprint,superscriptaddress]{revtex4-2}
\usepackage[english]{babel}
\usepackage[utf8x]{inputenc}
\usepackage[T1]{fontenc}
\usepackage{siunitx}
\usepackage{xcolor}
\usepackage{xspace}
\usepackage[version=4]{mhchem}
\usepackage{lineno}

\usepackage{placeins}
\usepackage{braket}
\usepackage{stackengine}
\usepackage{longtable}
\usepackage{booktabs}
\usepackage{natbib}
\usepackage{bibentry}
\usepackage{amsmath}
\usepackage{amssymb}
\usepackage{graphicx}
\usepackage[colorlinks=true, allcolors=blue,breaklinks=True]{hyperref}

\usepackage{ragged2e}

\usepackage{comment}
\usepackage{scalerel}

\begin{document}

\title{Recent advances in poled lithium niobate}


\author{Kibret A. Messalea}
\affiliation{%
Quantum Photonics Laboratory and Centre for Quantum Computation and Communication Technology, RMIT University, Melbourne, VIC 3001, Australia
}%

\author{Tim Weiss}
\affiliation{%
Quantum Photonics Laboratory and Centre for Quantum Computation and Communication Technology, RMIT University, Melbourne, VIC 3001, Australia
}%

\author{Yang Yang}
\affiliation{%
Quantum Photonics Laboratory and Centre for Quantum Computation and Communication Technology, RMIT University, Melbourne, VIC 3001, Australia
}%

\author{Hamed Arianfard}
\affiliation{%
Quantum Photonics Laboratory and Centre for Quantum Computation and Communication Technology, RMIT University, Melbourne, VIC 3001, Australia
}%

\author{Brett C. Johnson}
\email{brett.johnson2@rmit.edu.au}
\affiliation{School of Science, RMIT University, Melbourne, VIC 3001, Australia}

\author{Alberto Peruzzo}
\email{alberto.peruzzo@rmit.edu.au}
\affiliation{%
Quantum Photonics Laboratory and Centre for Quantum Computation and Communication Technology, RMIT University, Melbourne, VIC 3000, Australia
}%
\affiliation{%
Quandela, Massy, France}%


\begin{abstract}

Lithium niobate is a versatile material for both classical and quantum photonics, recognized for its outstanding electro-optic and nonlinear optical properties. Through a process known as poling, periodic ferroelectric crystal domains can be engineered to enable quasi-phase-matched frequency conversion, efficient modulation, and the generation of quantum light sources. The emergence of lithium niobate on insulator technology has further enhanced its suitability for scalable integrated photonics, offering ultra-low optical losses and strong light confinement while retaining the material’s inherent advantages. Here, the techniques used to fabricate and characterize periodically poled lithium niobate are reviewed. Key developments are discussed, offering insights into the future of domain engineering of lithium niobate.

\end{abstract}
\maketitle

\tableofcontents

\section{Introduction}
\label{sec:introduction}

Lithium Niobate (LN) represents a material platform central to a broad range of current and emerging optical technologies. It exhibits strong optical-nonlinear and electro-optic interaction and benefits from well-established device fabrication protocols, making it a scalable platform for integrated photonic devices~\cite{ZhuLoncar:21}. Applications leveraging its nonlinear interaction alone span from optical frequency conversion in the terahertz range~\cite{Kumar:90,Fejer:94} to the generation of non-classical light, including single photons, entangled photon pairs, and squeezed light, which is particularly difficult to produce by other means~\cite{ZhongPan:20,AasiZweizig:13,GiustinaZeilinger:15,NehraMarandi:22,KashiwazakiFurusawa:20}. Key to LN's appeal is its capability to implement so-called quasi-phase matching by locally inverting the material nonlinearlity---a technique referred to as poling. This enables efficient interaction for processes that can otherwise not be accessed, greatly broadening the platform's versatility and allowing for explicit engineering of the nonlinear interaction ~\cite{ArmstrongPershan:62,FrankenWard:63}. The fabrication of poled LN, however, is technically challenging and remains an active subject of research.

In the past decade, the availability of commercial lithium niobate on-insulator (LNOI) has greatly boosted both research interest and technological relevance of the field. The development of integrated nonlinear waveguides, in particular, has led to the development of devices with functionalities much beyond those realized in bulk crystals ~\cite{ArieVoloch:10,WeissPeruzzo:25,ZhangKrolikowski:21}. Because these technologies require high repeatability, fine domain sizes, and straightforward implementation and characterization of the poling process, significant efforts have been devoted to advancing its development.~\cite{chen:24}. Such efforts have, in turn, led to the development of new capabilities, resulting from, for example, the realization of sub-micron-size inverted domains~\cite{nagy:20,slautin:21} and the demonstration of two and three-dimensional poling structures~\cite{xu2022femtosecond}. 

This review explores the methods that have been developed to fabricate and characterise precision-engineered poled LN. We discuss the state of both poled bulk LN and poled LNOI substrates, and provide a comprehensive analysis of their applicability together with a discussion of their future within photonic technologies. 

\section{Background}
\label{section:background}

Lithium Niobate has long been recognized as a powerful platform in the field of photonics, and, due to its unique combination of nonlinear, electro-optic, and piezoelectric properties \cite{arizmendi2004photonic,qi2020integrated,boes2023lithium}, has been employed throughout a wide number of different applications. The following section will provide an overview of the platform and present brief introductions to the material properties, the mechanism underlying the poling process, and how it is applied throughout various photonic technologies.

\subsection{The properties of lithium niobate}

\begin{figure}
    \centering
    \includegraphics[width=0.5\columnwidth]{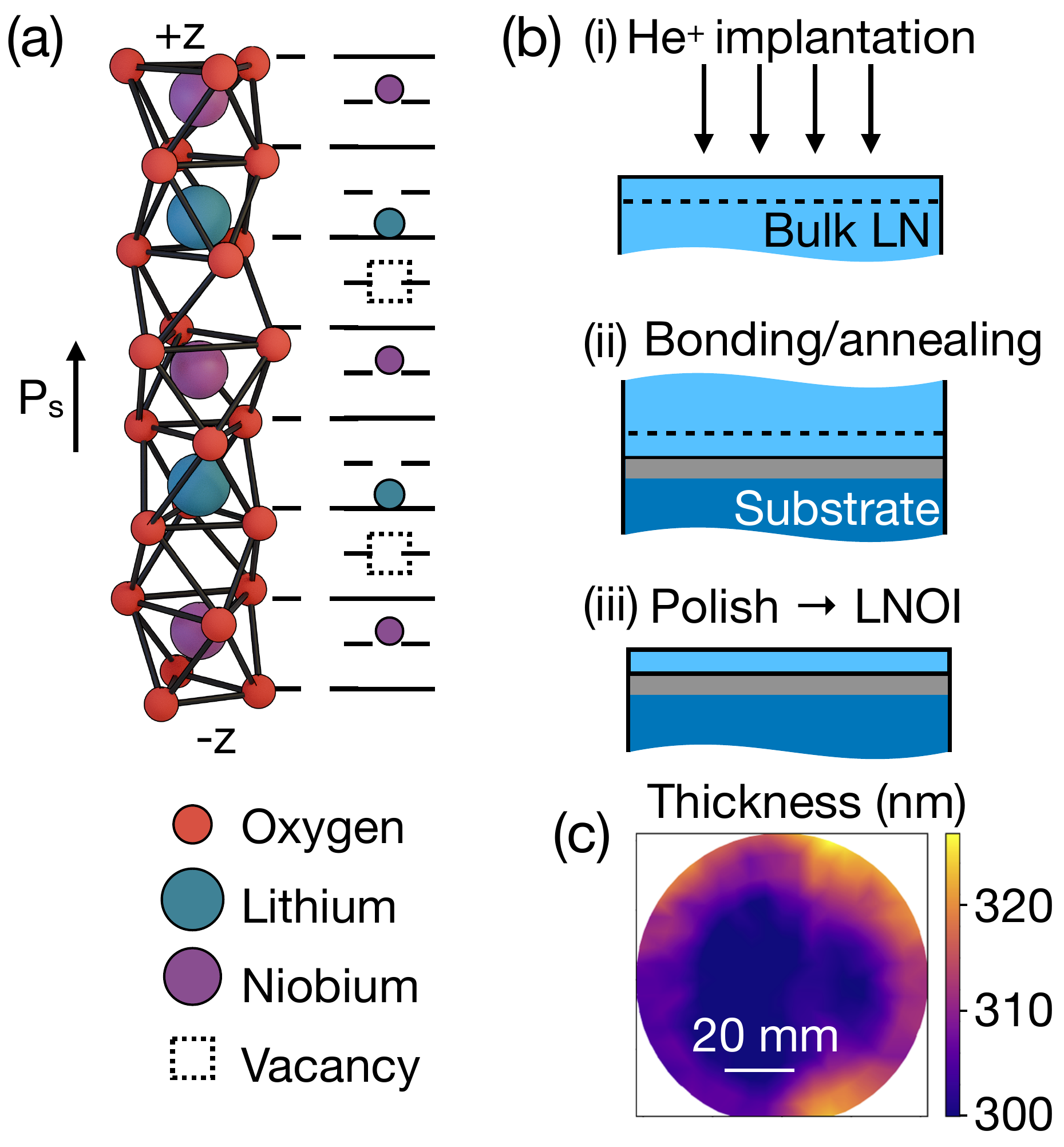}
    \caption{(a), Crystal structure unit cell of LN with spontaneous polarization image resulting from the displaced positions of the Li and Nb cations. (b) The main fabrication steps of LNOI, including (i)  He$^{+}$ ion implantation of bulk LN, (ii) bonding to a handle substrate and annealing to split the ion-implanted thin LN film, (iii) completed with surface polishing. (c), Measured LN thickness uniformity of a 4-inch wafer with 600 nm film thickness~\cite{luke:20}, reprinted with permission from ref.~\cite{luke:20} $\copyright$ Optical Society of America.}
    \label{fig:LNOI fab process}
\end{figure} 

Lithium Niobate is a synthetic material with a rhombohedral unit cell (space group R3c), as illustrated in Fig.~\ref{fig:LNOI fab process}(a). The unit cell consists of alternating layers of Lithium (Li$^+$), Niobium (Nb$^{5+}$), and Oxygen (O$^{2-}$) atoms. The crystal is composed of slightly distorted Oxygen octahedra surrounding both Lithium and Niobium atoms. The Lithium and Niobium atoms are slightly displaced along the z-axis, giving rise to a non-centrosymmetric structure. 

Due to its birefringence, LN crystals and wafers can be prepared in different crystallographic orientations, generally referred to as $X$-, $Y$-, and $Z$-cuts, wherein the respective label indicates the crystal axis which is perpendicular to the wafer surface (see Fig.~\ref{fig:LNOI fab process}(a) where the Z face is indicated). Which crystal cut is most suitable for a given application can be highly case-specific, and depend on, for example, the polarization of the incident light or the electrode configuration used for poling or implementation of electro-optic interaction. The choice of cut, accordingly, often represents an important consideration for both the fabrication and the operation of a given device.

Due to its lack of inversion symmetry, LN features both second-order $\chi^{(2)}$ and third-order $\chi^{(3)}$ nonlinear interaction, and, accordingly, can leverage both the associated three- and four-wave-mixing processes. The aforementioned displacement of the Lithium and Niobium atoms relative to the surrounding Oxygen further results in a unit cell with a permanent electric dipole~\cite{chen:22,sanchez:20}. This gives rise to a spontaneous polarization along the z-axis and is responsible for LN's ferroelectricity. The direction of the spontaneous polarization can, critically, be switched under a sufficiently strong external electric field---reorientating the Lithium and Niobium atoms and effectively flipping the optical axis---to create the locally constrained inverted domains at the core of the poling process~\cite{weis1985lithium}. 

The artificial structuring of the material with specific poling patterns can be used to engineer the second-order $\chi^{(2)}$ nonlinear interaction, which allows to tailor the respective processes to a number of vastly different applications. This is the central motivation behind the poling process and we will address its applications in more detail in the following section. 

                                              
Lithium Niobate crystal is fabricated using primarily the Czochralski method, producing commercially available wafers as large as 11.8 inches in diameter~\cite{weis1985lithium,chen:04}. Congruent LN (CLN) (a stable eutectic and slightly lithium-deficient composition, Li$_2$O:Nb$_2$O$_3$ = 48.38:51.62) is easier to grow than stoichiometric LN (SLN) but suffers from various limitations~\cite{sanchez:20,chen_LN:21}: Its high concentration of intrinsic defects makes it susceptible to photo-refractive and photo-chromic damage when exposed to high-power laser beams~\cite{kong:20,kong:12}, limiting its use in high-intensity or UV applications. Photo-refractive damage manifests as beam distortions due to refractive index variations caused by charge generation~\cite{kong:12}.
Photo-chromic effects result in absorption changes, such as green-induced infrared absorption~\cite{shin:04}.

Due to its resistance to photorefractive and photochromic effects, Magnesium-Oxide-doped LN (MgO:LN) has emerged as an often-preferred alternative to both CLN and SLN ~\cite{choubey:06}. Like in SLN, this resistance is rooted in the reduction of defect sites, most notably that of anti-site Niobium defects~\cite{grabmaier:86}. The coercive field necessary for domain inversion in MgO:LN and SLN crystals is four times lower~\cite{kurimura:01} than that of CLN, enabling the process in significantly thicker samples and enhancing the optical damage threshold~\cite{nakamura:02}.
Typically, all three material types are grown using the Czochralski method; MgO:LN currently represents the most common variant~\cite{choubey:06,grabmaier:86}.

The advent of ion slicing and wafer bonding technology (see in Fig.~\ref{fig:LNOI fab process}(b))~\cite{levy1998fabrication,rabiei2004optical} in the early 2000s enabled the production of high-quality thin-film LN wafers, which, over the last decade, have become commercially available. The LN thin-film is often further bonded to an insulator substrate, a combination referred to as Lithium Niobate on-insulator (LNOI). LNOI has become available in a variety of different thicknesses and crystal cuts and can be fabricated with high uniformity, a critical requirement when hosting nonlinear interactions. The buried oxide layer allows the confinement of light in the top layer~\cite{krasnokutska2018ultra}, reducing the device footprint, thereby allowing a greater density of devices on one compact platform~\cite{saravi2021lithium,vazimali:22}.

LN possesses a wide transparency window and high refractive indices (400~nm to 5~$\mu$m, $n_{o,e}$ > 2). Low-loss waveguides in bulk LN can be realized by either temperature induced in-diffusion of Titanium films deposited on the LN surface or by proton-ion exchange with Lithium ions using a hot acid bath. Although simple to implement, these techniques result in relatively large waveguides with low refractive index contrasts ($\sim$0.01 - 0.02) and comparably weak confinement. In contrast, LNOI waveguides can be fabricated using high-resolution lithography techniques and reactive ion etching. This allows to fabricate waveguides with high index contrast and wavelength scale confinement, enabling complex devices on the micrometer and nanometer scale.

\subsection{Poled lithium niobate} 

\begin{figure}
\centering
\includegraphics[width=14cm]{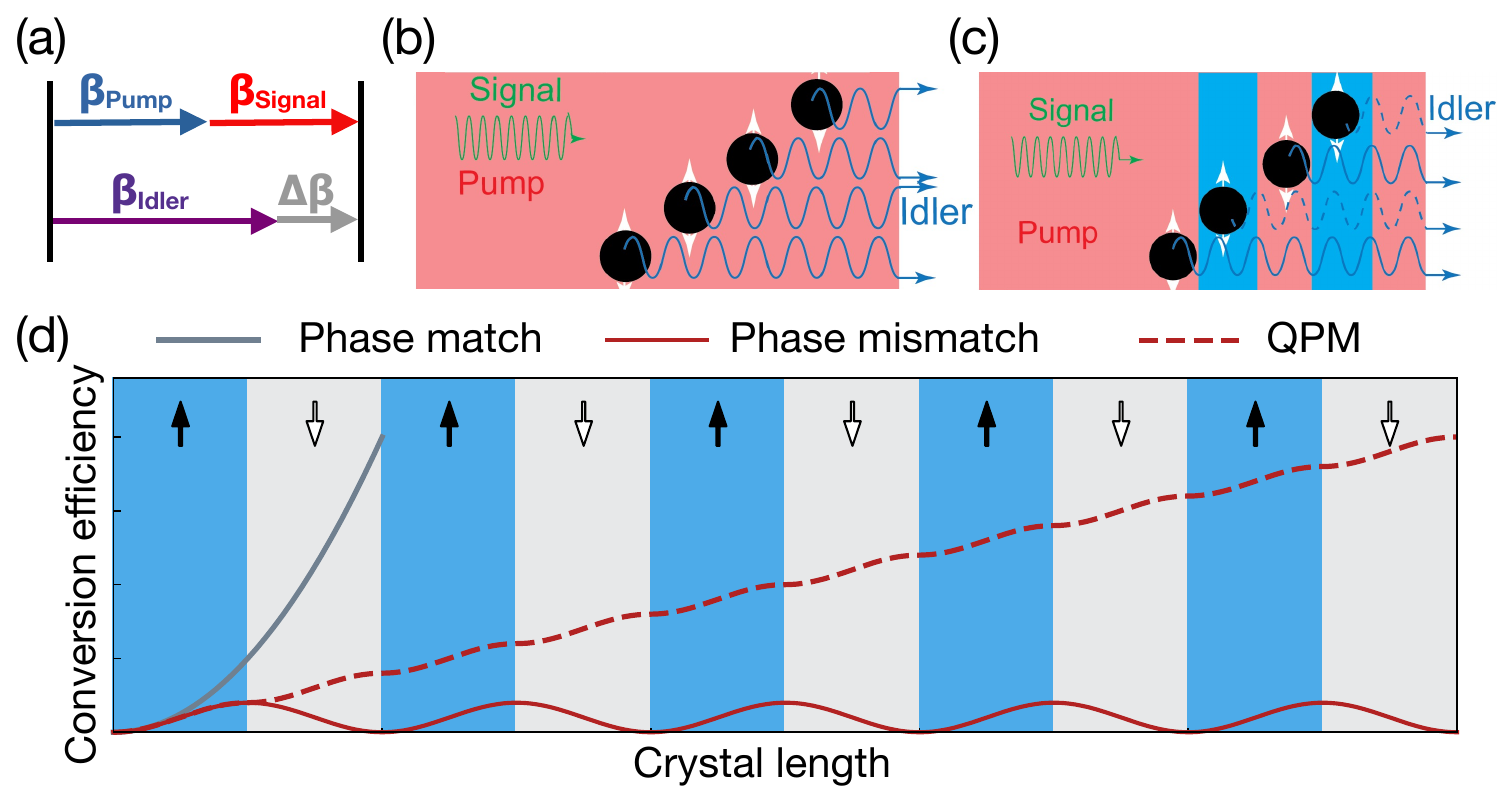}
\caption{\textbf{Quasi-phase matching (QPM) in second-order nonlinear processes.} (a) Phase-matching condition represented in terms of propagation constants ($\beta_{\text{pump}}$, $\beta_{\text{signal}}$, $\beta_{\text{idler}}$) and wave-vector mismatch $\Delta\beta$. (b) Due to material dispersion, conservation of momentum is generally not satisfied by default, resulting in a phase mismatch that prevents the efficient buildup of the generated radiation. (c) By periodically inverting the material nonlinearity, an artificial wavevector is introduced into the momentum conservation condition, enabling the coherent buildup of radiation from individual dipoles. (d) Conversion efficiency as a function of crystal length, comparing phase-mismatched, perfectly phase-matched (black solid line), and quasi-phase-matched interactions (red dashed line) through periodic domain inversion. The black and white arrows indicate the direction of the domains.}
\label{F3}
\end{figure}
           
Poled LN is formed by locally inverting the material’s nonlinear coefficient, achieved by reversing the sign of its second-order susceptibility through external stimuli that exceed the coercive field. Domain inversion is commonly achieved through electric-field poling, electron-beam or ion-beam writing, and light-based poling using UV or femtosecond lasers~\cite{ying:12,steigerwald:10}. In electric-field poling, a voltage is applied across the crystal to directly supply the coercive field required to invert the local polarization. In light-based approaches, absorbed photons generate electron-hole pairs and localized heating, temporarily reducing the spontaneous polarization. The resulting imbalance between depolarization and screening fields creates a net field that can exceed the coercive threshold, inverting the polarization. The exact dynamics of the domain inversion process are often influenced by localized factors, such as defects, charge diffusion, thermal effects, or the vector sum of fields from photo-induced contributions.

By artificially patterning the crystal with alternating segments of inverted and un-inverted nonlinear susceptibility of well-defined length, it becomes possible to engineer the nonlinear interactions of the interacting fields as they propagate through the crystal. The availability of precise, localized domain inversion is hereby critical, and can vastly extend the range and controllability of the nonlinear interaction~\cite{HumFejer:07,WeissPeruzzo:25,WeissPeruzzo:25(2)}.

The mechanism for this engineerability is rooted in the conservation laws at the core of the three-wave mixing process. Momentum conservation, in particular, determines how the different fields build up during propagation---a process which naturally, due to material dispersion, occurs in an incoherent manner in the vast majority of cases. Periodic inversion of the material nonlinearity can be used to introduce an artificial, effective wave vector, allowing to compensate the momentum-mismatch of almost any interaction and enabling the fields to build up coherently. This technique, commonly referred to as quasi-phase-matching (QPM), periodically restores constructive interference along the crystal, enabling effective buildup of the fields generated by the respective conversion process. Fig. \ref{F3} illustrates the principle of the QPM technique and the associated coherent buildup of the interacting fields.

Beyond enabling efficient conversion, structuring of the nonlinearity can represent a powerful degree of freedom for tailoring the properties of the generated light~\cite{HumFejer:07,HuZhu:13}. Using appropriately designed poling structures, one can selectively phase-match multiple different nonlinear processes, control the bandwidth of the output, or shape the spectral and temporal correlations of individual photons~\cite{AnsariSilberhorn:18}. Advanced designs exploit nonuniform or aperiodic poling to enhance specific spectral features or produce engineered entanglement structures \cite{WeissPeruzzo:25}. Over the last decade, such advanced QPM techniques have transformed LN into a versatile platform for both efficient classical frequency conversion and precisely controlled sources of quantum light, some major examples of which are explored in the next section.

\subsection{Major applications}

\begin{figure}
    \label{fig:applications}
    \centering
    \includegraphics[width=0.6\columnwidth]{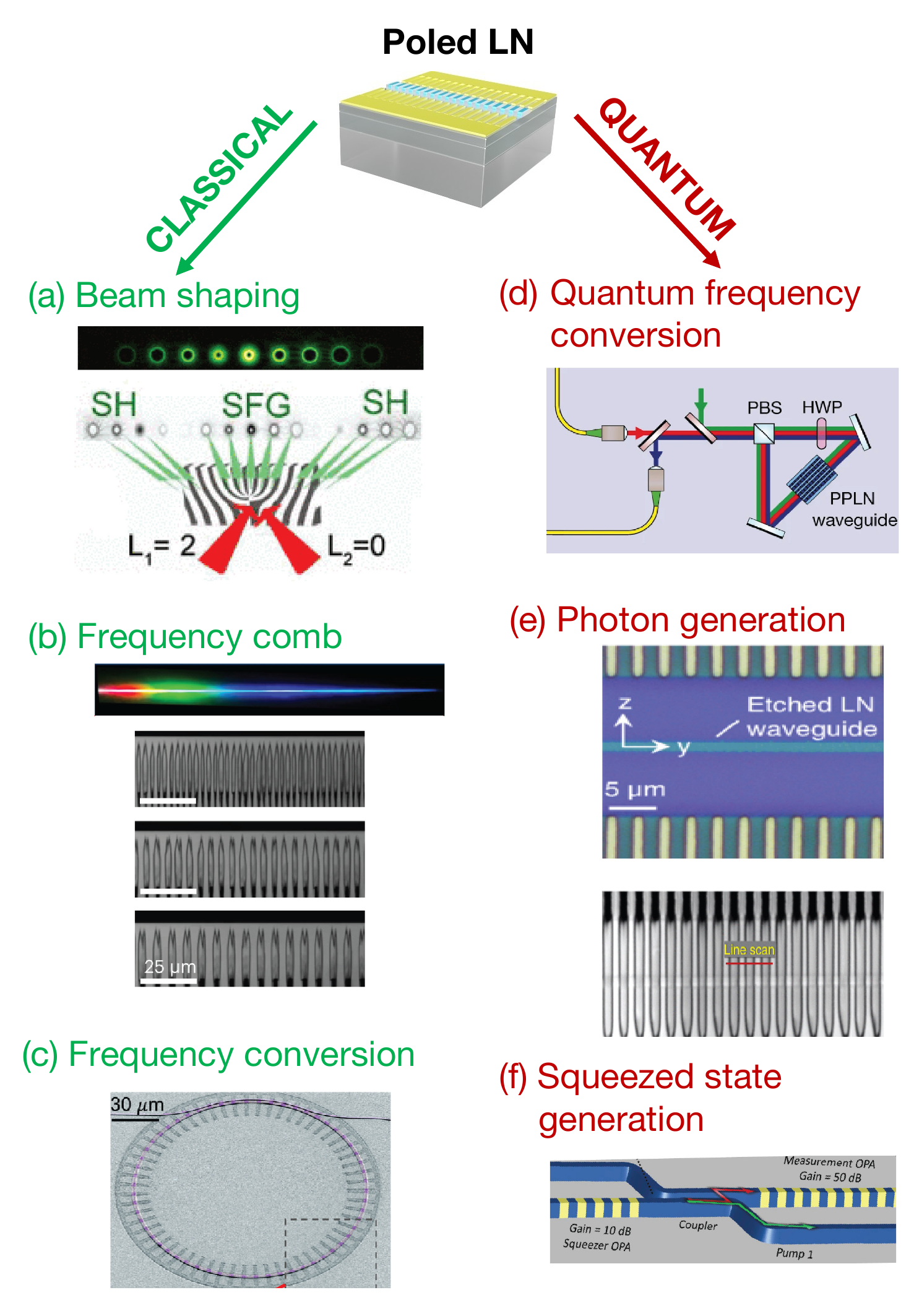}
    \caption{\textbf{Major technologies in classical and quantum applications of PPLN.} Schematic illustration of the applicability of structured LN crystals across photonic technologies with exemplary demonstrations from the recent literature, featuring (a) beam shaping techniques~\cite{BlochArie:12}, (b) supercontinuum frequency comb~\cite{WuDiddams:24}, (c) highly efficient SHG~\cite{LuTang:19}, (d) a memory-to-telecom interconnect~\cite{VanLeentWinfurter:22}, (e) photon-pair generation~\cite{ZhaoMookherjea:20} and (f) squeezed state generation and measurement~\cite{NehraMarandi:22}.}
    \label{F2}
\end{figure} 

LN has long served as a key platform for optical modulators, frequency conversion, and surface acoustic wave technologies. These devices are prominent in classical optical communication systems and signal processing technologies~\cite{wooten_2000_review,ZhuLoncar:21}. LN further enables the generation of light at wavelengths not readily available from standard laser sources and has been utilized to construct red, yellow, green, blue and ultra-violet lasers, by means of nonlinear harmonic generation~\cite{ThompsonMaleki:03,VanceMoosmueller:98,GeorgievTaylor:05,MeynFejer:97,PruneriHanna:9,BatchkoErman:99,MillerByer:97,LiuXu:01,WangKarlsson:98}. Varying the periodicity of the poling extends this capability, allowing for multicoloured lasing in a single device ~\cite{LiuXu:01(2),HuZhu:08,LeeChou:02,Capmany:01,FradkinArie:99,LiaoMing:03}. Further, more refined poling-structures have enabled broadband phase‑matching techniques~\cite{ChenLi:14,ChenLi:15,TehranchiKashyap:08,SuchowskiArie:14}, for example, in optical frequency comb generation~\cite{WuDiddams:24,KuesMorandotti:19} (Fig.~\ref{F2}(b)), and to support schemes that cascade multiple phase‑matching conditions to achieve white laser generation~\cite{HongLi:23,WeiChen:23}.

Patterns along two lateral directions of the crystal can enable nonlinear beam shaping techniques~\cite{ShapiraArie:15,HuZhu:20}. This has been used to generate vortex beams carrying orbital angular momentum~\cite{ShapiraArie:12,BlochArie:12}, and self-healing Airy beams~\cite{EllenbogenArie:09}.


Since the nonlinear interaction is at its core a quantum mechanical process, LN has been a key resource in the emerging field of photonic quantum technologies. So-called parametric down-conversion ~\cite{Couteau:18,JankowskiFejer:24}, a process in which pump photons are down-converted into pairs of single-photons, represents the primary source of quantum light. In the low-gain regime, this process generates correlated photon pairs and has become a cornerstone of quantum optics and quantum information science. Its versatility, operational simplicity, and stability at room temperature have enabled the realization of bright, high-purity single-photon sources~\cite{WangSun:21}, underpinning numerous landmark demonstrations~\cite{HongMandel:87,GiustinaZeilinger:15,YinLi:20,ZhongPan:20}. In the high-gain regime, parametric down-conversion forms the basis of optical parametric oscillators and amplifiers, and is the standard route to generating squeezed states of light, which play a critical role in quantum metrology and continuous variable information processing.

Advanced domain engineering techniques enable precise tailoring of the spectral, temporal, and entanglement properties of photons generated via the down-conversion process~\cite{WeissPeruzzo:25(2),AnsariSilberhorn:18}. Such techniques have been employed to generate optimized resources crucial for quantum communication, metrology, and multiphoton interference experiments, including spectrally pure single photons~\cite{GraffittiBranczyk:18,PickstonFedrizzi:21}, photons entangled in polarization~\cite{Sun:20,KuoNam:20} and time–frequency degrees of freedom~\cite{GraffittiFedrizzi:20,ShukhinEisenberg:23}, and broadband, ultrashort biphotons~\cite{SensarnHarris:10,ChekhovaPrudkovskii:18}.

The aforementioned classical frequency conversion capabilities can also be applied at the single photon level. Such quantum frequency conversion~\cite{Kumar:90,HuangKumar:92} has been employed to establish entanglement between telecommunication-band photons and single atoms~\cite {VanLeentWinfurter:20}, atomic ensembles~\cite {IkutaImoto:18,AlbrechtRiedmatten:14}, diamond spin qubits~\cite {DreauHanson:18}, and trapped ions~\cite {KrutyanskiyLanyon:19,BockEschner:18,WalkerKeller:18}. Demonstrations of entanglement between distant atomic systems have also been achieved with QFC~\cite{VanLeentWinfurter:22,LuoPan:22,YuPan:20,MaringRiedmatten:17}. Quantum frequency conversion at the single photon level has further found application in conjunction with existing technologies, extending the detection bandwidth of silicon avalanche-photon-diodes~\cite{WangPan:23,ZhengPan:20,PelcFejer:11,LangrockTakesue:05}, facilitating sensing schemes with single-photon resolution~\cite{DamPederson:12,ZhengZeng:23,WidarssonLaurell:22}, and enabling a number of further applications~\cite{KobayashiImoto:16,LiChen:19,QiChen:21,EcksteinSilberhorn:11,SensarnHarris:10}.

The versatility of nonlinear elements is further enhanced when combined with other photonic components. For example, cascading multiple nonlinear elements allows the construction of nonlinear interferometers~\cite{ChekhovaOu:16}, while integration with resonant or cavity structures enables the generation of quantum optical microcombs~\cite{KuesMorandotti:19} or optical parametric oscillators~\cite{LuOu:00,BruchTang:19}. We present an overview of the applications leveraging the nonlinear interaction of LN in Fig. \ref{fig:applications}

\section{Poling methods}\label{section:methods}

Ferroelectric domain engineering in LN can be achieved with various methods, such as electric field poling (EFP)~\cite{yamada1993first}, Ultraviolet (UV) laser direct writing~\cite{muir2008direct}, infrared (IR) femtosecond pulsed laser~\cite{xu2022femtosecond}, electron beam (EB)~\cite{restoin:00} and focused ion beam (FIB)~\cite{li2005nano} methods. The following sections describe these methods in turn.

\subsection{Electric field poling}

\begin{figure*}
    \centering
    \includegraphics[width=0.5\columnwidth]{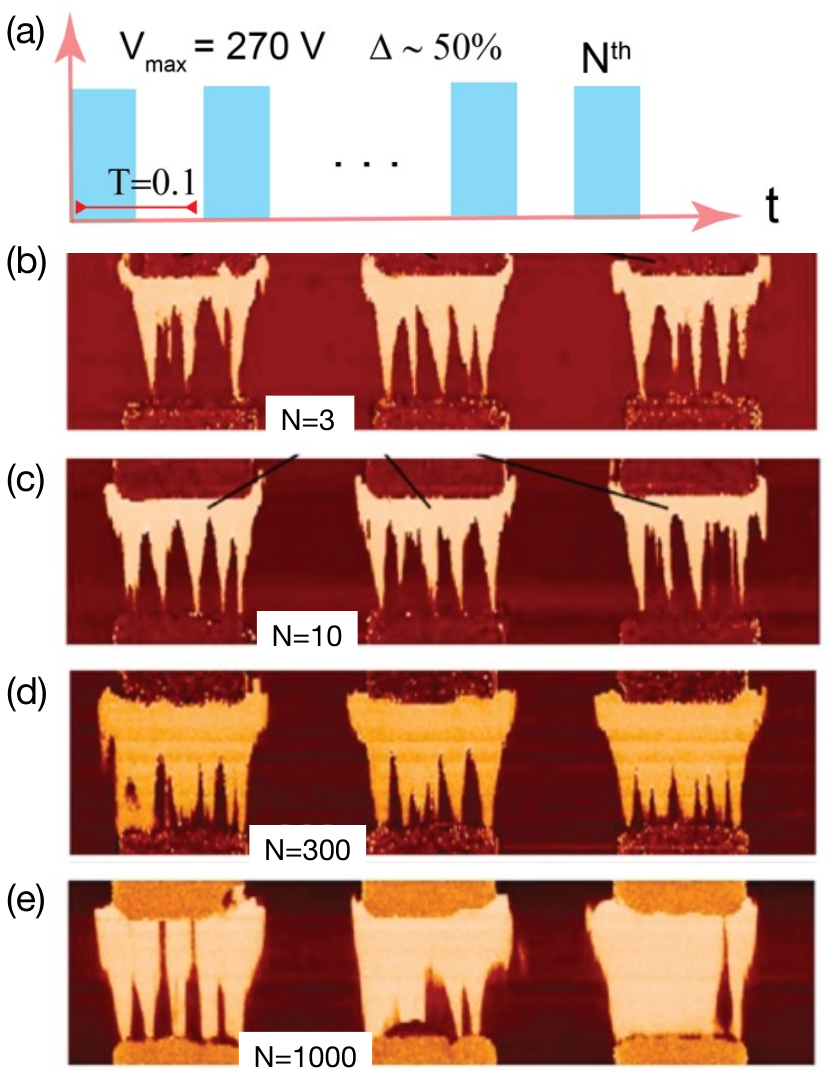}
    \caption{\textbf{EFP poling of X-cut LN and the effect of pulse number.} (a) Schematic of the EFP pulse sequence. (b)-(e) Piezoresponse force microscopy (PFM) images showing domain inversion after applying 3, 10, 300, and 1000 pulses, respectively. Panel (a)-(e) reprinted with permission from ref.~\cite{zhang:20} \copyright 2019 Elsevier.}
    \label{F5}
\end{figure*} 

EFP is the most widely used and well-established technique for producing PPLN. It is also the primary method employed for MgO-doped LN and thin-film LNOI substrates. Despite its broad adoption, EFP requires careful optimization of the applied voltage, multiple fabrication steps for electrode patterning, and often suffers from fabrication-induced defect production. 

When the applied electric field exceeds the coercive field of LN ($\sim 21$~kV/mm), domains invert by $180^\circ$, yielding the desired periodic poling pattern~\cite{Shur:15}. To achieve this, patterned electrodes are often fabricated on the +Z face of LN, aligned with the $d_{33}$ axis ($d_{33} \approx 28$~pm/V), which maximizes the nonlinear interaction strength. Once a high voltage is applied between the +Z and –Z surfaces, often under immersion in insulating oil to suppress breakdown, domain inversion initiates beneath the electrode edges and propagates through the bulk of the crystal.  

In the case of LNOI, X-cut substrates may be employed, where both the anode and cathode electrodes are lithographically defined as interdigitated structures. This geometry establishes a lateral electric field between adjacent electrode fingers, thereby driving domain inversion in the regions between the contacts.

The domain inversion process proceeds through three characteristic stages: nucleation, forward growth, and lateral stabilization. As shown in Fig.~\ref{F5}.1a–c, nucleation begins at electrode boundaries with the formation of nanoscale inverted domains. These domains extend into the crystal thickness as needle-like structures, which subsequently broaden laterally until the designed pattern is reached~\cite{Shur:15}. Ideally, the poling field is switched off at this stabilization stage, ensuring accurate domain morphology. However, abrupt removal of the field can cause repoling, while excessive poling duration can lead to overpoling, as illustrated in Fig.~\ref{F5}.1d–e. Both effects reduce the fidelity of the periodic structure.  

Although effective, EFP faces several intrinsic challenges. Domain broadening, driven by fringe electric fields, leads to deviations from the designed period~\cite{Shur:15,zhang:22}. Localized current leakage, often arising from material defects, can induce unintended domain formation~\cite{nataf:20}. Overpoling and repoling remain recurrent issues, particularly in thin-film geometries where precise control of the field is more difficult~\cite{zhang:22}. Furthermore, many emerging applications demand submicrometer-scale poling periods, which are exceedingly difficult to realize with conventional EFP without introducing significant defects.  

To address these issues, a range of improvements has been proposed. Spontaneous back-switching~\cite{batchko:99}, for example, improves uniformity by abruptly removing the poling voltage at carefully selected times, though its extension to LNOI remains limited. Bipolar pulse poling introduces alternating preconditioning pulses to control nucleation and stabilization, suppressing domain merging errors in short-period structures~\cite{nagy:20}. Alternatively, tailoring the number, shape, and duration of poling pulses has been shown to reduce poling errors. As demonstrated in Fig.~\ref{F5}.2a–f, increasing the number of pulses improves domain fidelity, though this approach primarily impacts the nucleation phase~\cite{stanicki:20,chang:16}. More advanced pulse shaping techniques, such as gradual voltage ramp-downs, mitigate abrupt repoling and enhance stabilization, as depicted in Fig.~\ref{F5}.3a–c~\cite{zhao:20}.  

Despite these refinements, EFP remains fundamentally constrained by its bulk nature, as the domains typically propagate through the entire crystal thickness. For many integrated photonics applications, however, it is sufficient—and often preferable—to confine poling to the near-surface region. Techniques that enable precise engineering of shallow, submicrometer-period domains (Fig.~\ref{F5}.4a–c) represent a promising direction for advancing LN-based nonlinear photonics. Control of maximum voltage and pulse sequences further suppresses domain broadening, as shown in Fig.~\ref{F5}.5a–j.  

Ongoing efforts focus on developing simplified, scalable processes capable of reliably producing short-period domain structures in thin films. Recent approaches demonstrate the feasibility of engineering nanoscale domain patterns with high fidelity (Fig.~\ref{F5}.6a–f), paving the way toward compact and efficient LN-based quantum and classical photonic devices.  

\subsection{Electron-beam poling}

\begin{figure}
    \centering
    \includegraphics[width=0.5\columnwidth]{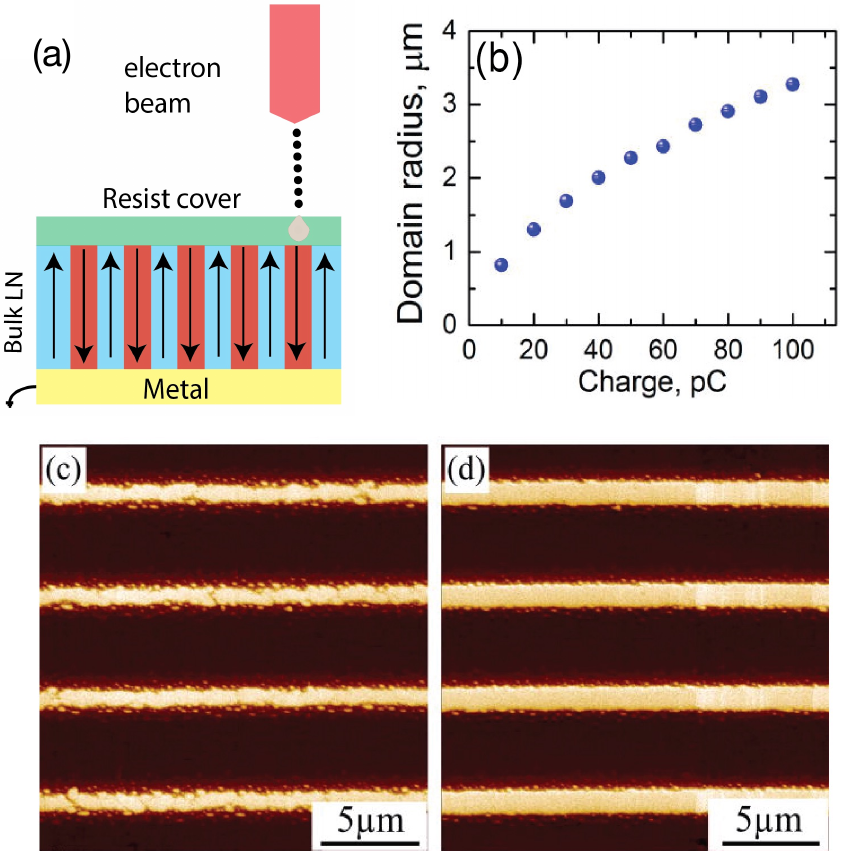}
    \caption{\textbf{Electron beam poling.} (a) Schematic of EB poling of bulk LN crystal~\cite{chezganov_e:16}. (b) Relationship between e-beam stripe exposure dose and inverted domain size. (c)\&(d) Piezo-force microscopy images of domain structures created using different exposure doses~\cite{chezganov_e:16}. Panel (b)-(d) reprinted with permission ref.~\cite{chezganov_e:16} \copyright 2016 AIP publishing.}
    \label{F7}
\end{figure} 

Electron-beam (EB) poling emerged in the early 1990s as an alternative method for domain inversion in LN~\cite{nutt:92}. In this approach, a focused electron beam, typically generated in a scanning electron microscope with energies of 10–30~keV, is directed onto the crystal surface. The incident electrons deposit charge in a near-surface region, thereby generating a localized electric field that can exceed the coercive field of LN and initiate domain reversal. At these energies, electrons penetrate several hundred nanometers to a few micrometers into the crystal, which aids in coupling the induced field more effectively into the bulk and promotes domain wall propagation beyond the initial nucleation region.  

Early demonstrations combined EB irradiation with an externally applied bias field to partially suppress the spontaneous polarization and thereby reduce the required electron dose~\cite{Lim1989}. Under these conditions, domain seeding could be achieved with very low EB exposure, while the applied bias field assisted subsequent domain growth. The technique has since been extended to a wider range of LN materials: in 2006, EB poling was successfully demonstrated in MgO-doped bulk crystals~\cite{li:06}, and more recently in 2017, quasi-phase-matched (QPM) gratings were realized in Ti-indiffused waveguides~\cite{chezganov_e:16}. To date, however, EB poling has not yet been applied to LNOI, the leading platform for integrated photonics.  

Compared with conventional EF poling, EB poling offers distinct advantages for fabricating engineered domain structures, particularly when feature sizes below the micrometer scale are required~\cite{chezganov_e:16,chezganov_e:17,savelyev:21}. However, the approach is limited by inhomogeneous space-charge distribution, which arises from the lateral spreading of incident electrons during exposure~\cite{chezganov_e:15}. Two strategies have been proposed to mitigate this effect. First, the deposition of thin dielectric layers, such as photoresists or electron-beam resists, can improve charge localization and thereby enhance the fidelity of the inverted domain pattern~\cite{chezganov_e:15,chezganov_e:16}. Second, tuning the electron beam energy provides a means to minimize space-charge effects by controlling the penetration depth and spatial distribution of deposited electrons~\cite{chezganov_e:17,vlasov:18,kokhanchik:17}.  

In practice, EB poling is most effective on the $-Z$ face of LN substrates, with typical exposure doses exceeding 1000~$\mu$C/$\mu$m$^2$~\cite{kokhanchik:17}. The method is compatible with standard electron-beam lithography tools and can be integrated into existing photonic device fabrication workflows without disrupting subsequent processing steps. Importantly, EB poling can be performed either before or after the fabrication of waveguides or other circuit components, providing flexibility in process integration.  

While EB poling has been convincingly demonstrated in bulk LN, including MgO-doped crystals (see Fig.~\ref{F7}), several limitations remain. These include the long processing times associated with the high exposure doses, as well as the restriction of the technique to the $-Z$ crystal surface~\cite{kokhanchik:22}. Nevertheless, the method offers several compelling advantages: (i) high-resolution domain patterning down to nanometer scales, (ii) the ability to realize submicron domain features, (iii) compatibility with non-planar or uneven surfaces, (iv) seamless integration with existing photonic nanofabrication facilities, (v) minimal crystal damage compared to focused ion beam (FIB) methods, and (vi) reduced risk of surface contamination or impurity incorporation. Together, these attributes position EB poling as a powerful tool for advanced domain engineering in LN, particularly where nanoscale precision and integration with photonic circuitry are required.

\subsection{Focused ion beam} 

\begin{figure}
    \centering
    \includegraphics[width=0.9\columnwidth]{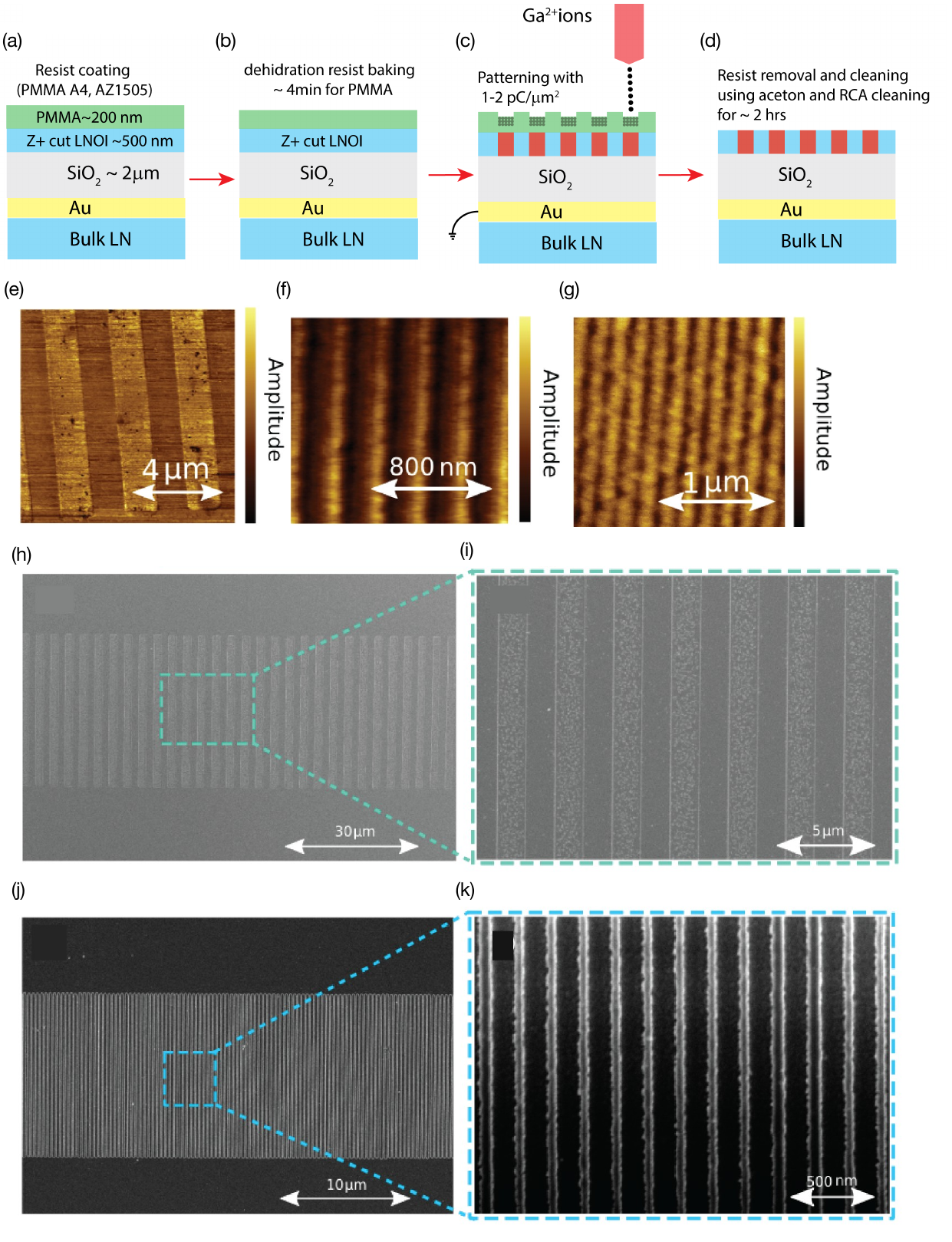}
    \caption{\textbf{Focused ion beam poling.} (a)–(d) Process flow: (a) deposition of a PMMA or photoresist layer to prevent Ga ion implantation and surface damage; (b) resist baking to improve resolution; (c) FIB patterning; (d) surface cleaning using RCA. (e)–(g) PFM images of domains poled in z-cut LNOI~\cite{krasnokutska:21}; (h)–(k) HF etching and SEM imaging of the poled region.}
    \label{F6}
\end{figure}

The use of FIB methods for domain inversion in LN was first reported in the early 2000s. As shown in Fig.~\ref{F6}, the technique has since been extended from bulk LN crystals in 2005~\cite{li:05}, to MgO-doped bulk substrates in 2017~\cite{chezganov:17}, and more recently to thin-film z-cut LNOI in 2021~\cite{krasnokutska:21}. 

FIB poling has emerged as a compelling alternative to electric-field poling (EFP), particularly for LNOI, owing to its recent commercial availability~\cite{chezganov:20,chezganov:23}. The technique enables submicron domain engineering, as ferroelectric crystals exhibit far less backscattering under focused ion irradiation than under electron-beam exposure~\cite{li2005nano}. A resist layer is typically deposited on the LN surface before irradiation to mitigate gallium ion implantation into the LN, which would generate defects and amorphisation~\cite{chezganov:17,krasnokutska:21,li:06}.  

The poling process proceeds as follows: (i) a focused $Ga^{2+}$ ion beam is directed onto the Z$^+$ surface of the crystal; (ii) ions accumulate at the surface, creating a dense sheet of positive charge; and (iii) once the resulting electric field surpasses the coercive field, the local spontaneous polarization of LN is inverted. Schematics of this process are illustrated in Fig.~\ref{F6}.  

Compared to EB poling, FIB poling achieves significantly faster processing times due to a roughly three orders of magnitude reduction in required charge dose. This advantage makes it attractive for scalable fabrication and high-throughput device production. Moreover, the technique is less material-dependent than EB or laser poling, broadening its applicability to a wider range of ferroelectrics.  

A further strength of FIB poling is its ability to define domains with submicron resolution~\cite{krasnokutska:21}, a capability that is essential for quasi-phase-matching at short wavelengths and for processes such as backward SHG. In this regime, implicit cavity effects can dramatically enhance nonlinear interactions, reducing or eliminating the need for engineered photonic resonators such as Bragg gratings or microring cavities. In addition, FIB methods permit domain writing on uneven or structured surfaces~\cite{krasnokutska:21}, thereby expanding the design space for integrated photonic devices.  

Finally, FIB poling offers flexibility in the fabrication workflow. Domains can be patterned either before or after waveguide fabrication~\cite{pashnina:23}. Post-fabrication poling, in particular, avoids scattering losses introduced by differential etching of the +Z and –Z crystal faces~\cite{chezganov:23}, thereby enabling higher-performance photonic devices.  

These advances establish FIB poling as a powerful tool for realizing complex, high-resolution domain patterns in LNOI. Its compatibility with established semiconductor processing platforms highlights its potential for next-generation integrated nonlinear and quantum photonic devices.  

FIB poling thus provides a versatile means of creating domain patterns in LN and LNOI with submicron resolution. While early demonstrations have confirmed the feasibility of the approach, integration with more complex photonic architectures is yet to be explored. These open questions suggest that further development of FIB poling could yield new opportunities for domain engineering in integrated nonlinear photonics.

\subsection{Laser direct write poling}

Laser poling is an all-optical method for ferroelectric domain inversion that uses focused laser irradiation to locally modify the spontaneous polarization of LN. Depending on the laser wavelength, domain inversion can be driven either by surface-absorbed ultraviolet (UV) light or by near-infrared (NIR) femtosecond pulses. In both cases, multiphoton absorption at the focal point generates a local temperature rise and associated thermoelectric field, which can exceed the coercive field and reorient the ferroelectric domains. A key distinction is that UV light is strongly absorbed near the surface, restricting domain inversion to shallow depths, while tightly focused femtosecond NIR pulses penetrate deeper into the crystal, enabling true three-dimensional (3D) domain engineering. This capacity for volumetric structuring is the hallmark of laser-based poling compared with electric-field or charge-based approaches.  

\subsubsection{UV laser poling}

UV laser irradiation can invert domains in LN through the generation of a space-charge electric field. This arises from three coupled effects~\cite{muir2008direct}. Namely, photoexcitation of electrons and holes, drift of these carriers under an applied electric field, and a reduction in spontaneous polarization at elevated temperatures.  

Since electrons and holes have unequal mobilities, the +Z and –Z faces of LN respond differently to UV exposure~\cite{muir2008direct,mailis:10}. UV irradiation can thus be used either to \emph{assist} electric-field poling or, under certain conditions, to directly invert domains without an external bias. If the UV wavelength is close to the LN bandgap, it increases resistance to poling, whereas longer wavelengths reduce the coercive field permanently~\cite{mailis:10,steigerwald:10}. Direct UV poling is possible on the –Z face, where the strongly absorbed light can create sufficient space-charge fields for domain inversion~\cite{sebastian:23,imbrock:18}.  

The depth of UV-induced domain inversion is typically limited to a few microns, with values of $\sim 100$~nm on the +Z face and up to $\sim 1~\mu$m on the –Z face~\cite{imbrock:18}. This shallow penetration makes UV poling especially relevant for thin-film and LNOI platforms, while also allowing extension to non-polar cuts. However, the inherently surface-confined nature of the process limits its applicability to bulk 3D domain engineering.

\subsubsection{Femtosecond laser poling}

Femtosecond laser poling exploits tightly focused, ultrafast NIR pulses to directly write ferroelectric domains with sub-100~nm resolution inside LN crystals~\cite{wang:23,yu:08,wang:24,zhang:21}. In contrast to conventional electric-field poling, which is restricted to two-dimensional domain inversion with micrometer-scale resolution, femtosecond laser writing enables complex three-dimensional domain architectures~\cite{xu2022femtosecond,zhang:20}.  

Recent work has demonstrated non-reciprocal femtosecond poling, where the directionality of the laser beam governs domain growth~\cite{xu2022femtosecond}. In this scheme, multiphoton absorption at the focal point generates a highly localized temperature gradient, producing a thermoelectric field that dominates domain inversion, with only a minor contribution from the pyroelectric effect. Using this approach, periodic domains with a lateral resolution as small as 30~nm have been achieved~\cite{imbrock:18}.  

\begin{figure}
    \centering
    \includegraphics[width=1\columnwidth]{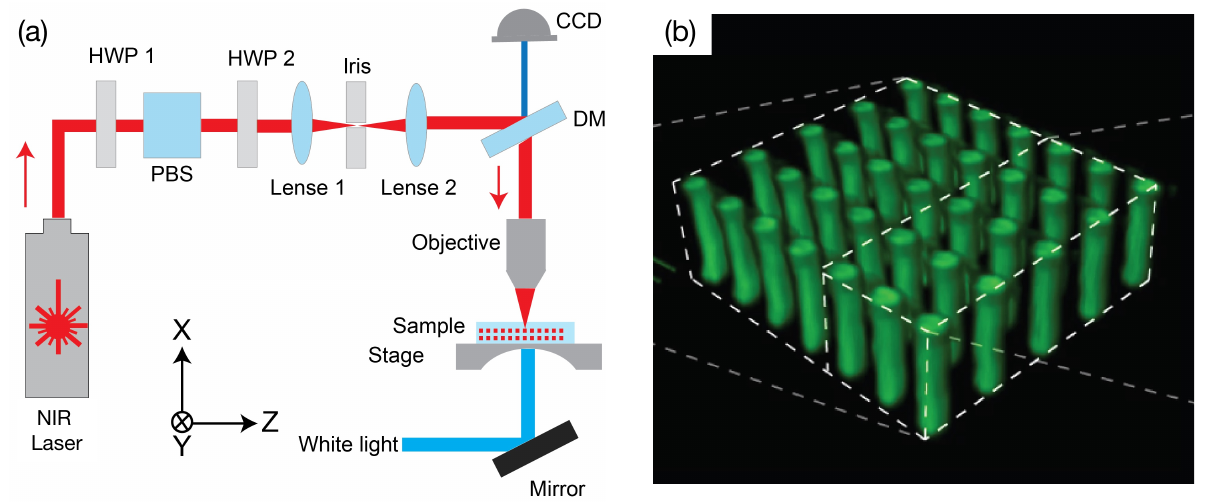}
    \caption{\textbf{Femtosecond laser poling.} (a) Schematic of the femtosecond laser direct-write setup~\cite{xu2022femtosecond}. (b) Cherenkov-type SH microscopy of 3D nanodomain structures~\cite{xu2022femtosecond}. Panel (b) reprinted with permission from ref.~\cite{xu2022femtosecond}, \copyright Springer Nature Ltd.}
    \label{fig:Femtosecond laser poling}
\end{figure}

Characterization of such nanodomains by piezoresponse force microscopy (PFM) and Cherenkov-type second-harmonic microscopy has confirmed true 3D structuring capabilities, as shown in Fig.~\ref{fig:Femtosecond laser poling}. Beyond domain engineering, femtosecond-poled LN crystals provide a platform for tailoring the spatial distribution of the second-order nonlinear susceptibility, enabling manipulation of nonlinear wavefronts, enhanced quasi-phase matching, and novel applications in entangled photon generation and quantum photonics.  

\subsection{Comparison} 

Over the past three decades, LN domain engineering has advanced considerably, producing increasingly efficient and accurate techniques. Among these, the electric field poling (EFP) method remains the most widely adopted due to its robustness and versatility. It applies to a wide range of geometries and has been extended to thin-film LN using electrode-free approaches such as piezoresponse force microscopy (PFM)~\cite{ma:23}. Nevertheless, EFP has inherent limitations: sub-micron periodicities are difficult to achieve because of uncontrolled domain broadening~\cite{stanicki:20}, and the process requires patterned electrodes and high-voltage insulators to suppress fringe fields~\cite{chang:16}. Despite these drawbacks, EFP continues to be the most reliable choice for many phase-matching applications, particularly in x-cut LNOI where lower voltages suffice~\cite{zhao:20,nagy:20}.

Alternative methods have emerged with higher resolution and accuracy. Electron beam (EB) and focused ion beam (FIB) poling provide superior spatial control, enabling sub-micron periodic structures with reduced error~\cite{chezganov:17,krasnokutska:21}. EB poling has matured from requiring external field assistance to a standalone technique applicable to bulk LN and integrated waveguides. Its main limitations are the slow writing speed and confinement to the -Z crystal face, which complicates integration with subsequent electron-beam lithography due to potential re-poling from high-dose exposures. FIB poling, in contrast, benefits from the smaller interaction volume of ions, requiring lower doses and minimizing back-scattering. This enables efficient, high-resolution poling with minimal risk of re-polarization, making it particularly attractive for short-pitch periodicities. However, FIB remains less established for LNOI substrates, though recent demonstrations indicate its potential applicability~\cite{krasnokutska:21,chezganov:23}.

UV-assisted poling exploits LN’s absorption edge, where exposure on the -Z face suppresses inversion but on the +Z face promotes it, owing to differences in carrier mobility~\cite{steigerwald:10,ying:12}. While straightforward and low-cost, UV poling has not yet been tested on LNOI, limiting its present relevance. 

Femtosecond laser poling has recently shown exceptional promise, particularly in three-dimensional domain engineering~\cite{xu2022femtosecond,chen:23}. This method achieves nanometer-scale resolution and allows for dynamic writing and erasure of domain structures without mechanical repositioning~\cite{imbrock:18,dai:23}. Although not yet demonstrated on LNOI, its versatility and precision suggest strong potential for future integrated photonics applications.

In summary, Table~\ref{tab:Methods} highlights the trade-offs among the various poling methods. EFP remains the most practical and established technique, particularly for large-area periodic structures. EB and FIB provide unmatched accuracy at the sub-micron scale, with FIB offering faster processing and reduced re-poling risks. UV and femtosecond laser methods present alternative routes with unique advantages in simplicity and 3D structuring, respectively, but their integration with LNOI is still unproven. Continued research will be required to fully exploit these emerging approaches and determine their roles alongside established methods in next-generation LN photonic devices. 

\begin{table*}
\caption{Comparison of poling methods }
  \label{tab:Methods}
  \centering
\begin{tabular}{@{} l *4c @{}}
\toprule
 \multicolumn{1}{c}{Methods}    & Complexity  & Compatibility  & Sub-micron poling  & PPLN accuracy \\ 
\midrule
 EF  & Medium & All & No & Low \\ 
 EB  & Low & -Z face & Yes & High\\ 
 FIB   & Low & +Z face & Yes & High \\
 UV   & Low & -Z face & Yes & Medium \\ 
 Femtosecond   & Low & All & Yes & High \\ \bottomrule
 \end{tabular}
\end{table*}

\section{Poling characterization}\label{section:charaterization}

The performance of PPLN is closely linked to the quality of the poling process. Critical parameters include the uniformity of the poling pattern relative to the design, the consistency of the duty cycle (ratio of poled to unpoled regions), and the depth of domain inversion, particularly within the optical waveguide region~\cite{nagy:20,reitzig:21,ruesing:19}.

Accurate characterization of these parameters is essential to achieve high-fidelity periodic poling, minimizing deviations in period, duty cycle, and poling depth~\cite{ruesing:19,stanicki:20,reitzig:21}. Both non-destructive and destructive techniques have been employed to evaluate poling quality~\cite{brockmeier:21,shur:10}. Prominent methods include microscopy-based approaches such as piezoresponse force microscopy (PFM)~\cite{cherifi:21}, confocal Raman spectroscopy~\cite{rusing:18}, and confocal SHG ~\cite{spychala:20,hegarty:22}, as well as etching-based techniques, including hydrofluoric acid wet etching and inductively coupled plasma (ICP) dry etching~\cite{bullen:16}.

Poling errors commonly manifest as deviations from the intended pattern, including variations in period and duty cycle, incomplete domain inversion, overpoling, underpoling, and insufficient poling depth~\cite{reitzig:21}. These errors can reduce the efficiency of quasi-phase-matched (QPM) nonlinear interactions, highlighting the need for precise assessment of periodic poling quality, particularly in real-time during device fabrication.

Table~\ref{tab:Compare_characterization} summarizes several widely used domain visualization techniques, comparing their resolution, 3D imaging capability, destructiveness, and compatibility with LNOI substrates. While all poling methods are susceptible to some degree of domain randomness, electric field poling in particular is prone to irregularities that can impact phase-matching performance.

The following sections provide an overview of these characterization methods and discuss how they reveal different types of poling errors.

\begin{table*}
\caption{Domain visualization methods comparison}
  \label{tab:Compare_characterization}
  \centering
\begin{tabular}{@{} l *4c @{}}
\toprule
 \multicolumn{1}{c}{Methods}    & Destructive  & High-resolution  & 3D Visualization  &  Compatibility with LNOI  \\ 
\midrule
 PFM  & Yes & Yes(10 nm\cite{reitzig:21}) & No & Yes  \\ 
 SH Confocal microscopy  & No & Yes(200 nm\cite{reitzig:21}) & No & Yes \\ 
 Cherenkov SH microscopy   & No & Yes(50 nm\cite{kampfe:14}) & Yes & No \\
 Confocal Raman microscopy   & No & No(250 nm\cite{shur:10}) & Yes & Yes \\ 
 HF wet etching with SEM   & Yes & Yes(1 nm\cite{shur:10}) & Yes & Yes \\ \bottomrule
 \end{tabular}
\end{table*}

\subsection{Piezo-force microscopy}

\begin{figure}
    \centering
    \includegraphics[width=1\columnwidth]{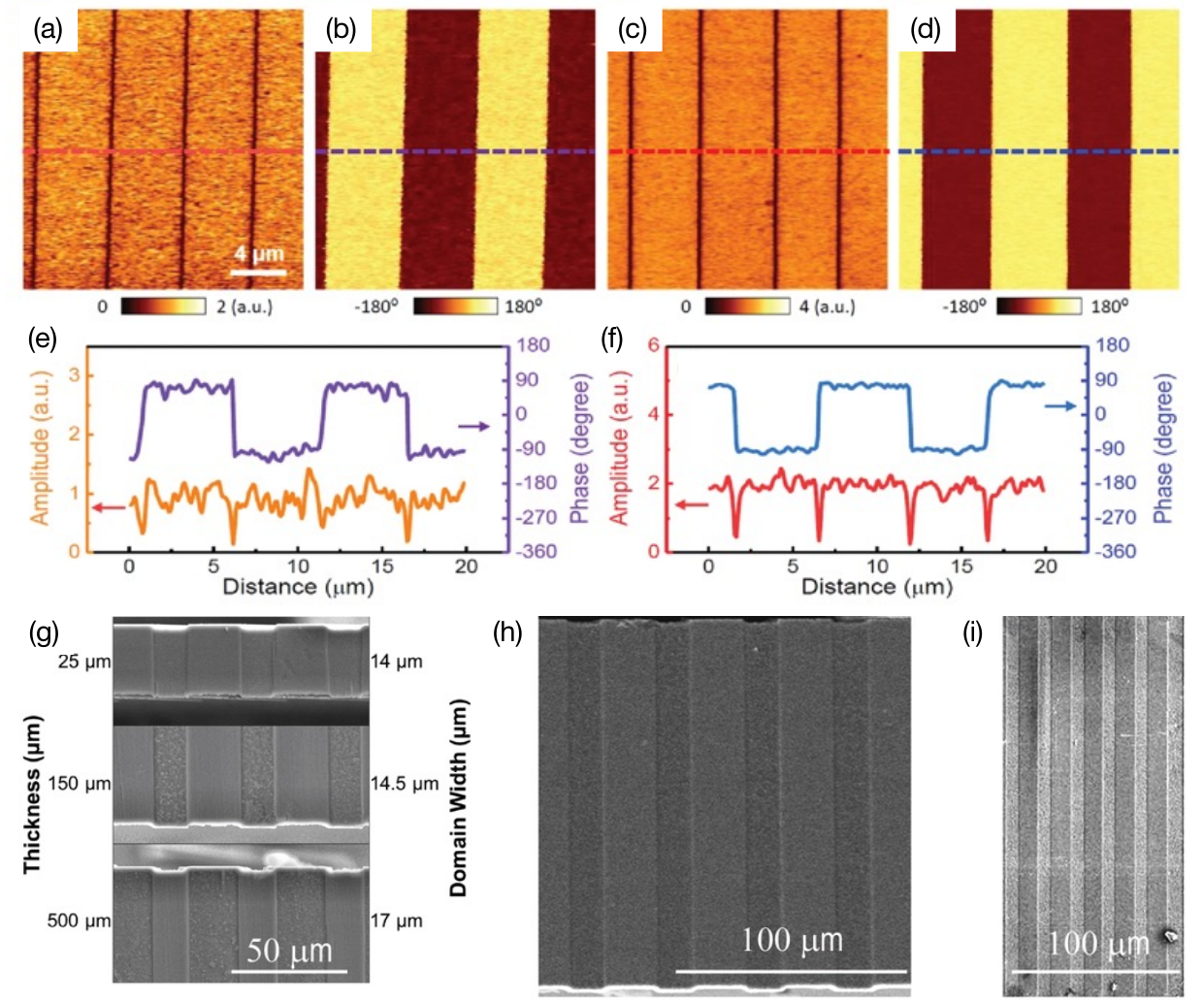}
    \caption{\textbf{Characterization of poling quality using PFM and SEM.} (a)-(f) PFM characterization resulting images~\cite{kim:17}.  (g)-(i) Hydrofluoric acid etched periodic poled LN and resulting SEM images~\cite{bullen:16}. Panel (a)-(f) reprinted with permission from ref.~\cite{kim:17} and permitted under the Creative Commons license. Panel (g)-(i) reprinted with permission from ref.~\cite{bullen:16} \copyright 2016 Elsevier.}
    \label{fig:poling characterizaion figure}
\end{figure}

Piezo-response force microscopy (PFM) is a non-destructive, contact-mode scanning probe technique widely used to study ferroelectric domains in LN~\cite{roeper:24,reitzig:21}. In PFM, an AC voltage is applied between a conductive atomic force microscopy (AFM) tip and the sample, which is typically mounted on a conductive stage. The oscillating electric field induces a local piezoelectric strain in the crystal, causing periodic surface deformations proportional to the corresponding tensor component, $d_{ij}$. These deformations are transmitted to the cantilever and detected as tip deflections, which are then demodulated by a lock-in amplifier. By monitoring both amplitude and phase, PFM reconstructs the ferroelectric domain orientation with nano-meter-scale lateral resolution, as shown in Fig.~\ref{fig:poling characterizaion figure}(a)–(f).  

The phase signal directly distinguishes between domains of opposite polarization, while the amplitude reflects the magnitude of the electromechanical response. Analysis of amplitude profiles across domain walls can provide information about wall depth and sharpness, while combined amplitude and phase maps allow quantitative evaluation of duty cycle deviations in periodically poled structures~\cite{roeper:24,chuhigh:18,slautin:21}. The acquisition times can be relatively slow and scan areas are typically below $100\times 100 \; \mu$m. Furthermore, the requirement for contact-mode operation leads to tip wear, possible surface damage, and variable signal quality depending on tip conductivity. 

Although chip-scale integration does not seem practical at present, selected area analysis may be useful for fabrication process development. Furthermore, its ability to provide depth-sensitive information is particularly advantageous for thin-film LNOI characterization~\cite{roeper:24}, which is at a spatial resolution much greater than standard optical characterisation methdos can achieve.  

\subsection{Scanning electron microscopy based on hydrofluoric acid etching}

Hydrofluoric acid (HF) etching exploits the different etch rates between inverted and non-inverted ferroelectric domains in LN~\cite{mailis:14,sones:02}. This approach provides a rapid, preliminary assessment of the poling process by revealing variations in domain structure. Following etching, scanning electron microscopy (SEM) is typically used to image the resulting surface microstructures, offering high-resolution insight into surface morphology and potential non-uniformities introduced during poling~\cite{stanicki:20}. Representative HF-etched PPLN and corresponding SEM images are shown in Fig.~\ref{fig:poling characterizaion figure} (g)-(i).

Despite its utility, HF etching has several limitations. It is primarily applicable to bulk LN, as etching can damage underlying SiO$_2$ layers and bonding adhesives in thin-film LNOI structures. The method is inherently destructive, preventing repeated measurements on the same sample. Moreover, its depth resolution is limited: features below approximately 1~$\mu$m are difficult to resolve, making HF etching less effective for thin-film LN (~500~nm or less). Nonetheless, it remains valuable for assessing key poling parameters such as duty cycle, period, and surface uniformity in bulk LN samples.

\subsection{Second harmonic generation microscopy}

SHG microscopy enables detailed mapping of the second harmonic intensity across ferroelectric materials~\cite{spychala:20,nataf:20}. The SHG signal depends on several factors, including the polarization and intensity of the incident laser, the local material properties represented by the nonlinear susceptibility tensor, and the relative phase of interacting light waves~\cite{reitzig:21}. Variations in signal intensity can indicate changes in the local susceptibility at domain walls, either due to modifications in the tensor magnitude or the appearance of new tensor components~\cite{zhao:20}.  

SHG microscopy is a non-destructive, high-contrast technique for evaluating the quality of poling in LN crystals. It provides rapid feedback on domain structure and poling fidelity, allowing visualization of domain walls in situ during or after poling. The technique is typically performed in a back-reflection geometry using femtosecond optical pulses, which makes it highly sensitive to local changes in crystal symmetry and structure.

For thin-film LNOI, SHG signals can be affected by interfacial reflections and resonant enhancements, effects that are negligible in bulk LN. Nevertheless, SHG microscopy can be adapted for thin-film characterization with minimal modifications. Fig.~\ref{F10} (a)\&(d) illustrate the SHG setup and representative images of poled LN.

In practice, a femtosecond pulsed laser in the NIR range is focused onto the sample through an objective lens. SH light generated in the focal region is collected in back-scattering mode via the same objective. Reflected fundamental light is suppressed using a dichroic beam splitter and a bandpass filter, and the filtered SH signal is detected by a single-photon avalanche diode (SPAD) through a single-mode optical fiber, which also serves as a confocal pinhole. This arrangement allows high-resolution imaging of domain walls and rapid assessment of poling quality.

\subsection{Cherenkov second harmonic generation microscopy}
Cherenkov SHG (C-SHG) microscopy is a powerful technique for three-dimensional visualization of spatial variations in the nonlinear susceptibility, $\chi^{(2)}$, of ferroelectric crystals such as LN~\cite{roppo:13,sheng:10}. As illustrated in Fig.~\ref{F10} (b)\&(e), C-SHG arises when the second harmonic signal is emitted at a specific angle determined by a longitudinal phase-matching condition~\cite{kampfe:14}. Signal enhancement occurs at domain wall interfaces, allowing high-contrast imaging of domain boundaries~\cite{sheng:10,roppo:13}. By scanning a tightly focused laser across the sample, C-SHG enables full three-dimensional reconstruction of the ferroelectric domain structure~\cite{xu2022femtosecond}.

C-SHG offers several advantages: it is non-destructive, provides sub-100~nm resolution of domain boundaries, and is particularly effective for characterizing domains produced via femtosecond laser poling~\cite{xu2022femtosecond}. While applicable to multiple poling methods, its use in LNOI-based domain imaging is currently limited. A notable limitation is the requirement for high-intensity laser beams, which can, in fact, induce localized heating and potentially alter the domain structure.

\subsection{Confocal Raman microscopy}

The Raman effect is the inelastic scattering of light by phonons, producing frequency shifts that reveal the vibrational modes of a crystal. These modes are highly sensitive to crystal symmetry, phase, and strain. Because the Raman tensor depends on the orientation of the local crystal axes with respect to the incident and scattered fields, the relative intensity and polarization of Raman peaks change with the ferroelectric domain orientation. This makes Raman spectroscopy a powerful probe of domain structures, since regions of reversed polarization exhibit complementary spectral signatures~\cite{rusing:18,rubio:15,nataf:20}.  

Confocal Raman microscopy (CRM) combines this spectroscopic sensitivity with the three-dimensional imaging capabilities of confocal microscopy. A tightly focused laser beam excites a microscopic volume within the crystal, and the backscattered light is passed through a filter which suppresses the  Rayleigh scattered light before entering a spectrometer and detector (CCD array). The confocal pinhole and a high numerical aperture (NA) objective provide sub-micron spatial resolution, typically about 1~$\mu$m laterally and 1-2~$\mu$m in depth, depending on the wavelength and refractive index~\cite{shur:10,rusing:18,chezganov:23}. Automated sample stages enable raster scanning to generate high-resolution spectral maps over large areas~\cite{zelenovskiy:10}.  

\begin{figure}
    \centering
    \includegraphics[width=1\columnwidth]{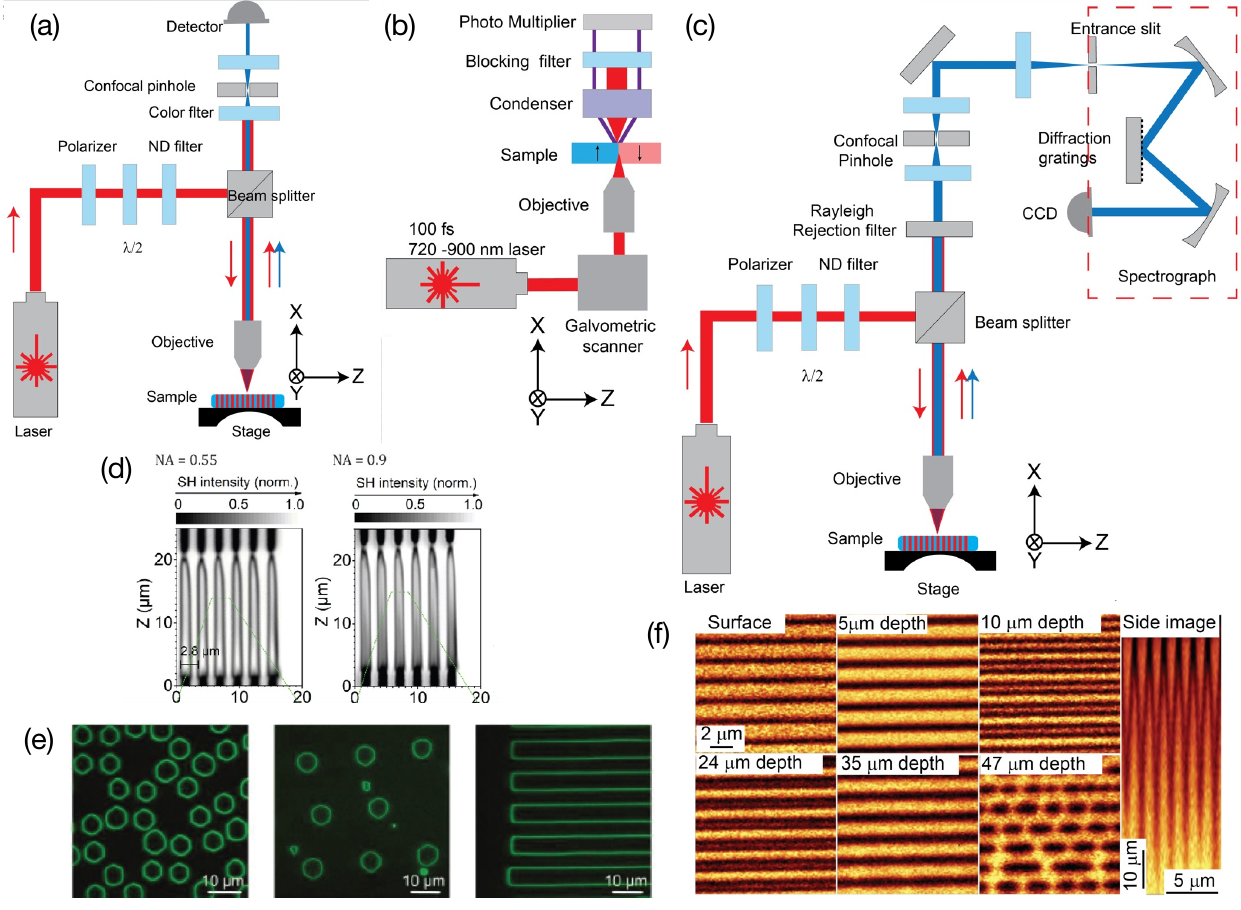}
    \caption{\textbf{Characterization of poling quality using optical methods.} (a)\&(d) SHG confocal schematics and SHG images using confocal microscopy~\cite{ruesing:19}. (b)\&(e) Cherenkov SHG microscopy setup schematics~\cite{kampfe:14} and resulting 2D domain imaging~\cite{sheng:10}. (c)\&(f) Confocal Raman Microscopy and resulting images~\cite{chezganov:23}. Panel (d) reprinted with permission from ref.~\cite{ruesing:19} \copyright 2019 AIP publishing. Panel (e) reprinted with permission from ref.~\cite{sheng:10} \copyright Optical Society of America. Panel (f) reprinted with permission from ref.~\cite{chezganov:23} \copyright 2023 Elsevier.}
    \label{F10}
\end{figure}

In LN, the intensity ratio of $A_1$ and $E$ phonon modes can be used to distinguish domains of opposite orientation~\cite{ya:14}. By analyzing such contrast across the scanned region, CRM provides not only images of the domain pattern but also quantitative information about local strain, defects, and domain wall properties. As shown in Fig.~\ref{F10}(c)\&(f), this nondestructive and label-free technique yields structural maps with high spatial resolution, offering detailed insight into the quality and uniformity of engineered ferroelectric domains.  

The Raman effect is relatively weak so that lasers of appreciable power are required to measure signals within a reasonable time. Care must therefore be taken to avoid laser heating which may alter the local domain structure. Deconvoling sprectral changes due to strain and those caused by domain inversion can be challenging. This is best alleviated with complementary measurements or polarization-resolved analysis.

\section{Summary and outlook}\label{section:conclusion}
Over the past decades, LN domain engineering has undergone significant evolution, underpinning many advanced applications in photonics and quantum technologies. EFP remains the most widely used method due to its effectiveness in producing high-quality, periodically and aperiodically poled LN structures. Nonetheless, challenges such as domain broadening, repoling, and overpoling persist, particularly as device miniaturization and submicron precision become increasingly important. To address these limitations, alternative approaches—including FIB poling, EB poling, and laser-based poling—have emerged. FIB poling offers high-resolution submicron domain patterns, EB poling enables precise domain inversion with minimal surface damage, and laser-based methods (UV and femtosecond) allow controlled domain inversion through localized heating or photo-excitation.

Advances in characterization have complemented these poling developments. Techniques such as PFM, SHG microscopy, Cerenkov SHG, confocal Raman microscopy, and HF etching provide critical feedback on domain quality, duty cycle, and depth, enabling the detection of deviations from design and supporting the optimization of nanoscale and integrated photonic devices.

The shift from bulk LN to thin-film LNOI substrates represents a transformative step, offering compactness, enhanced optical confinement, and compatibility with standard nanofabrication processes. While LNOI introduces challenges for domain inversion due to its thin-film structure, ongoing research is advancing methods that improve precision, reduce voltage requirements, and enhance scalability. FIB poling, in particular, shows strong potential for adaptation to LNOI and integration with established semiconductor workflows, facilitating complex photonic circuit fabrication. Hybrid strategies combining electric field and laser-based poling may further mitigate individual limitations, enabling refined micro- and nanoscale domain engineering.

Looking forward, continued development of poling techniques tailored for LNOI, alongside improvements in characterization and fabrication, will be critical to realizing scalable, high-performance photonic systems. These advancements are expected to underpin a new generation of compact, efficient, and versatile devices with wide-ranging implications across nonlinear optics, quantum technologies, and beyond.

\section*{Acknowledgments}
AP acknowledges an RMIT University Vice Chancellor’s Senior Research Fellowship and a Google Faculty Research Award. This work was supported by the Australian Government through the Australian Research Council under the Centre of Excellence scheme (No: CE170100012). 

\section*{Conflict of interest}
The authors declare no conflict of interest.

\clearpage
\bibliography{references}
\end{document}